\pdfoutput=1

\PassOptionsToPackage{table}{xcolor}
\documentclass[sigconf, screen]{acmart}

\renewcommand\footnotetextcopyrightpermission[1]{} 

\makeatletter
\renewcommand\@formatdoi[1]{\ignorespaces}
\makeatother

\AtBeginDocument{
  }

\setcopyright{none} 

\usepackage{listings}
\usepackage{enumitem}
\usepackage{float}
\usepackage{caption}
\usepackage{subcaption}
\usepackage{multirow}
\usepackage[most]{tcolorbox}

\tcbset{width=\columnwidth,boxrule=0pt,colback=red,arc=0pt,auto outer arc,left=0pt,right=0pt,boxsep=5pt}

\definecolor{customcolor}{HTML}{9494b8}

\newtcolorbox[auto counter]{ReqBox}{
    borderline west={0pt}{0pt}{white},
    colback=lightgray!30!white,breakable}

\newtcolorbox[auto counter]{ObsBox}{
    borderline west={0pt}{0pt}{white},
    colback=lightgray!30!white,breakable}

\newcommand{\req}[1]{\begin{ReqBox} #1 \end{ReqBox}}

\newcommand{\obs}[1]{\begin{ObsBox} \textbf{Observation~\thetcbcounter:} #1 \end{ObsBox}}

\definecolor{codegreen}{rgb}{0,0.6,0}
\definecolor{codegray}{rgb}{0.5,0.5,0.5}
\definecolor{codepurple}{rgb}{0.58,0,0.82}
\definecolor{backcolour}{rgb}{0.95,0.95,0.92}

\lstdefinestyle{mystyle}{
    commentstyle=\color{codegreen},
    keywordstyle=\color{magenta},
    numberstyle=\tiny\color{codegray},
    stringstyle=\color{codepurple},
    basicstyle=\ttfamily\footnotesize,
    breakatwhitespace=false,         
    breaklines=true,                 
    captionpos=b,                    
    keepspaces=true,                 
    numbers=left,                    
    numbersep=5pt,                  
    showspaces=false,                
    showstringspaces=false,
    showtabs=false,                  
    tabsize=2,
    frame=lines
}

\setlist[itemize]{leftmargin=*}
\setlist[enumerate]{leftmargin=*}

\author{Jiakun Yan}
\email{jiakuny3@illinois.edu}
\orcid{0000-0002-6917-5525}
\affiliation{
  \institution{University of Illinois Urbana-Champaign}
 \city{Urbana}
 \state{IL}
 \country{USA}
}
\author{Hartmut Kaiser}
\email{hkaiser@cct.lsu.edu}
\orcid{0000-0002-8712-2806}
\affiliation{
  \institution{Louisiana State University}
 \city{Baton Rouge}
 \state{LA}
 \country{USA}
}
\author{Marc Snir}
\email{snir@illinois.edu}
\orcid{0000-0002-3504-2468}
\affiliation{
  \institution{University of Illinois Urbana-Champaign}
 \city{Urbana}
 \state{IL}
 \country{USA}
}

\begin{document}

\title{Understanding the Communication Needs of Asynchronous Many-Task Systems -- A Case Study of HPX+LCI}

\begin{abstract}
  Asynchronous Many-Task (AMT) systems offer a potential solution for efficiently programming complicated scientific applications on extreme-scale heterogeneous architectures. However, they exhibit different communication needs from traditional bulk-synchronous parallel (BSP) applications, posing new challenges for underlying communication libraries. This work systematically studies the communication needs of AMTs and explores how communication libraries can be structured to better satisfy them through a case study of a real-world AMT system, HPX. 
  We first examine its communication stack layout and formalize the communication abstraction that underlying communication libraries need to support.
  We then analyze its current MPI backend (parcelport) and identify four categories of needs that are not typical in the BSP model and are not well covered by the MPI standard. To bridge these gaps, we design from the native network layer and incorporate various techniques, including one-sided communication, queue-based completion notification, explicit progressing, and different ways of resource contention mitigation, in a new parcelport with an experimental communication library, LCI. Overall, the resulting LCI parcelport outperforms the existing MPI parcelport with up to 50x in microbenchmarks and 2x in a real-world application. Using it as a testbed, we design LCI parcelport variants to quantify the performance contributions of each technique. This work combines conceptual analysis and experiment results to offer a practical guideline for the future development of communication libraries and AMT communication layers.
\end{abstract}


\maketitle
\pagestyle{plain} 

\section{Introduction}

\begin{table*}
    \centering
    \begin{tabular}{lllll}
    \toprule
     Category & Need & Reason & Major Technique Investigated & Benefit \\
     \midrule
\rowcolor{lightgray!20} Asynchrony &Efficient receive of & One-sided task invocation & Pre-posting receive and periodical testing & Baseline \\
\rowcolor{lightgray!20}            & unexpected messages && One-sided communication primitive & Low \\
     Concurrency & Efficient polling of
                        & Task oversubscription
                                & Request pool polling             & Baseline \\
                 & many pending operations &
                                & Queue-based completion mechanism & Medium \\
                              &&& Different queue implementation   & High\\
\rowcolor{lightgray!20} Multithreading   & Efficient multithreaded   & Multithreaded AMT         & Single device with coarse-grained blocking locks  & Baseline \\
\rowcolor{lightgray!20}             &communication              & runtime                   & Avoidance of coarse-grained blocking locks        & High \\
\rowcolor{lightgray!20}             &                           &                           & Avoidance of coarse-grained try locks             & Medium \\
\rowcolor{lightgray!20}             &                           &                           & Replication of devices                            & High \\
                        Progress    & Explicit progress engine  & Multiple AMT subsystems   & Implicit progress when testing                    & Baseline \\
                                    & invocation control        & to progress               & Explicit progress function                        & Medium \\
     \bottomrule
\end{tabular}
    \caption{High-level overview of AMT communication needs, underlying reasons, major techniques, and practical benefits discussed in this paper. Practical benefits represent the overall performance improvement of the corresponding technique over the "baseline" technique, estimated through HPX-level experiments. Complicated interactions between techniques are omitted here.}
    \label{table:requirement_overview}
\end{table*}

As high-performance computing enters the exascale era, architectures evolve toward increasing heterogeneity, rich on-node parallelism, and deep memory hierarchy~\cite{llnl_elcapitan, lanl_venado} while scientific applications employ more complicated and dynamic algorithms~\cite{hofmeyr2020metahipmer, GenomePaRSEC2024Ltaief}. The "bulk-synchronous parallel" (BSP) model, which assumes a fixed number of compute resources executing at the same speed with evenly distributed workloads, does not match such algorithms and does not map well onto extreme-scale heterogeneous architectures. These challenges have led to an increased interest in the \emph{Asynchronous Many-Task} (AMT) model~\cite{kulkarni2019comparative}. The AMT model expresses the application logic as a set of fine-grained sequential tasks and dependencies between them. An AMT runtime controls the scheduling of tasks, enforcing an execution order consistent with the dependencies. It manages the inter-task communications needed to satisfy these dependencies, thus removing the expensive global synchronization in the BSP model. In addition, AMT applications are usually over-decomposed (with more concurrent tasks than hardware threads) 
to increase communication overlaps and reduce the impact of the variability of the execution speed and workload~\cite{castillo2019optimizing}. During recent years, several large-scale AMT applications have been developed and showcased their programmability and performance advantages: Legate~\cite{bauer2019legate_numpy, Yadav2023legate_sparse} automatically parallelizes and scales popular Python packages such as Numpy and SciPy to distributed GPU clusters leveraging the Legion runtime~\cite{bauer2012LegionExpressingLocality}; Octo-Tiger~\cite{daiss2024octotiger} enables portable and scalable adaptive multi-physics simulation of stellar mergers using HPX~\cite{Kaiser2022HPX}; 
High-resolution Earth System Model simulations are scaled to 100's of  Petaflops~\cite{EarthSystemPaRSEC2024Abdulah} using the PaRSEC runtime~\cite{bosilca2013PaRSECExploitingHeterogeneitya}.

\begin{figure}[htbp]
    \centering
    \begin{subfigure}{0.49\linewidth}
        \includegraphics[width=\linewidth]{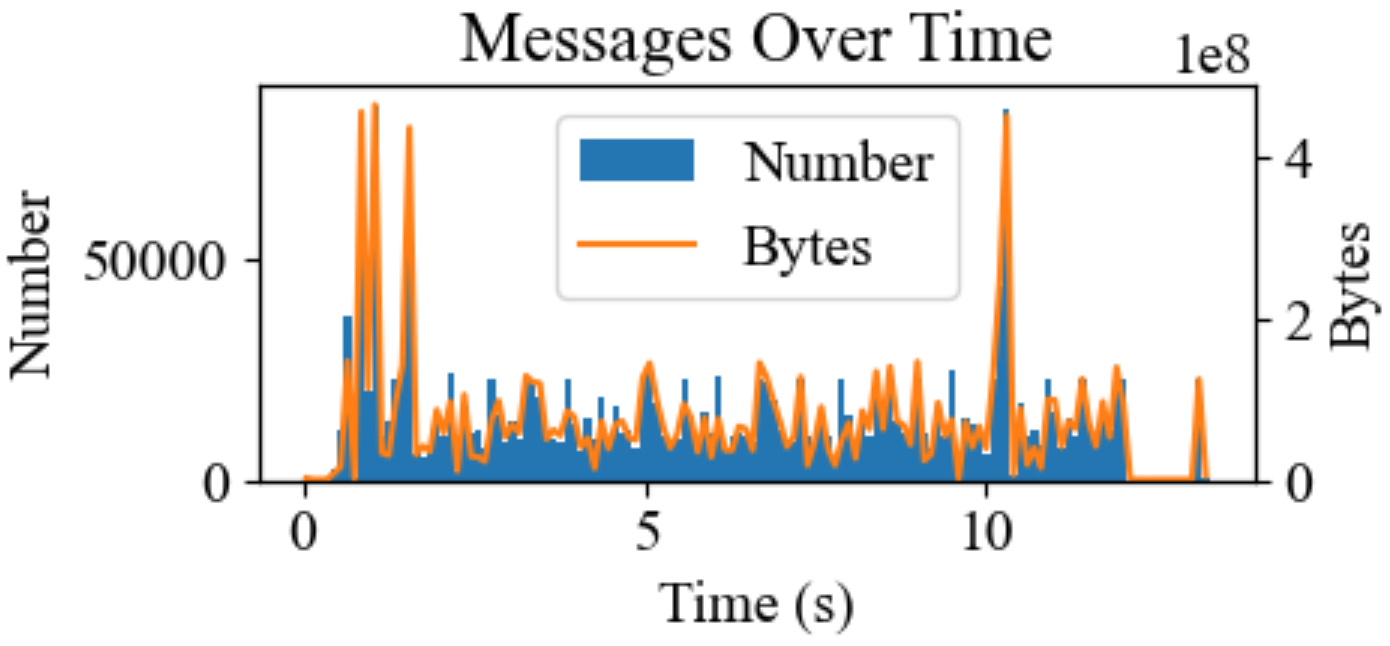}
        \caption{Messages over time.}
        \label{fig:octotiger-size_trend}
    \end{subfigure}
    \begin{subfigure}{0.49\linewidth}
        \includegraphics[width=\linewidth]{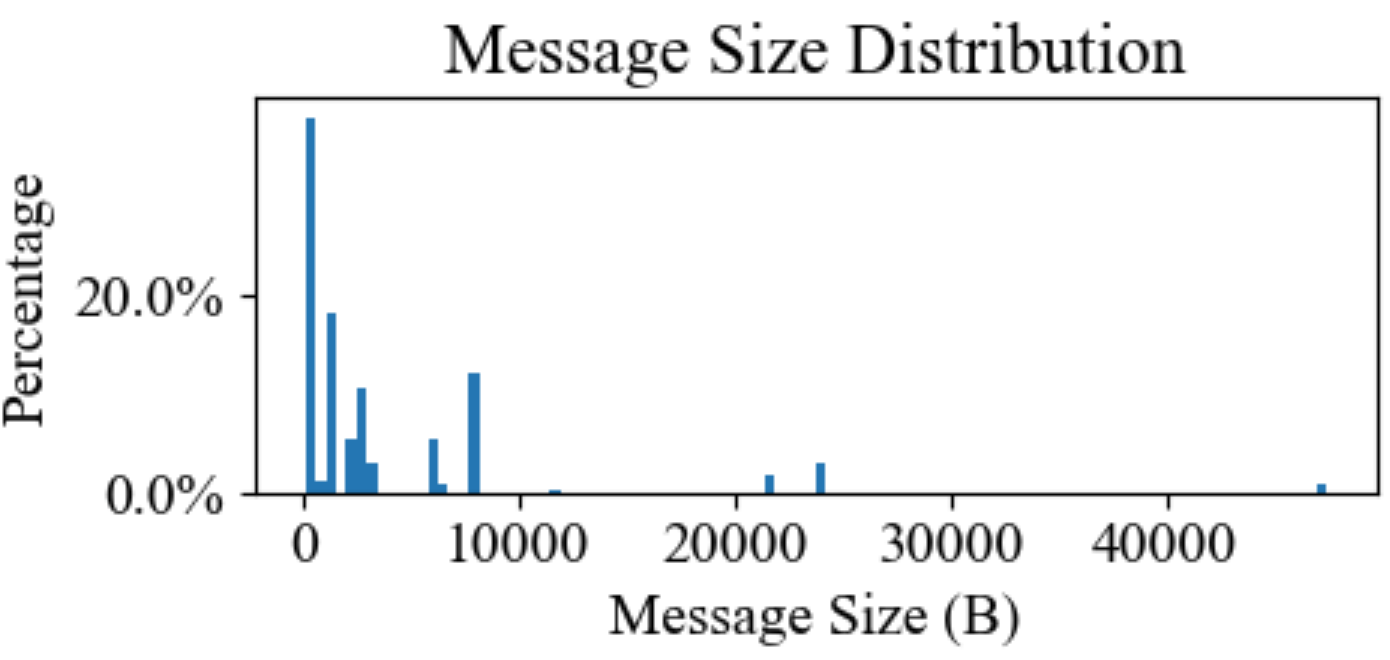}
        \caption{message size distribution.}
        \label{fig:octotiger-size_dist}
    \end{subfigure}
    \caption{Communication profile of the "rotating star" scenario of Octo-Tiger running on top of HPX.} 
    \label{fig:octotiger-comm}
\end{figure}

However, when scaling to distributed clusters, AMTs exhibit different communication profiles compared to those of traditional BSP applications~\cite{treichler2014RealmEventbasedLowlevel,mor2023PaRSEC_LCI}: a large number of messages can be logically produced and consumed concurrently and asynchronously by all the executing threads; communication requests include a mix of short control messages consumed by the runtime and longer data transfer messages consumed by compute tasks; most communications are point-to-point; there is no clear separation of computation and communication phases, and communication frequency and volume can vary across space and time. Figure~\ref{fig:octotiger-comm} showcases some of these characteristics with the communication profile of Octo-Tiger: communication happens all the time with high frequency and varying volume (Figure~\ref{fig:octotiger-size_trend}); and communication is dominated by small messages, with occasional large messages (Figure~\ref{fig:octotiger-size_dist}). These characteristics are usually not efficiently covered and sufficiently tested in mainstream communication libraries (i.e. MPI)~\cite{zambre2022LessonsLearnedMPI, Patinyasakdikul2019openmpi_mt, Bernholdt2020MPIsurvey, mpich_vci_build_issue, mpich_hang_expanse} and can result in reduced performance and increased programming complexity in AMTs. Conversely, there is very limited documentation about the abstraction and design of AMTs' communication stacks in existing literature, and many design decisions are not systematically explored or evaluated (This is further discussed in Section~\ref{sec:related_work}). Therefore, communication library developers have limited guidance when designing new features to support AMTs better. This motivates our research: we seek to systematically understand the communication needs of AMTs and how communication libraries should be structured to satisfy AMTs' needs. We perform this research through an in-depth case study of an established AMT system, HPX. Table~\ref{table:requirement_overview} gives an overview of this work. 

The first part of this work is to understand the problem. Before this project, HPX used MPI as its main communication backend (which HPX calls \emph{parcelport}). We systematically studied HPX's communication stack layout and its usage of MPI. We summarize HPX's communication abstraction into the \emph{HPX parcelport abstraction}, which formalizes the interface a communication library needs to implement to support HPX. We then analyze its current MPI parcelport and identify several inefficiencies. We find they are usually rooted in the interface and/or implementation limitations of existing communication libraries, causing AMT communications to be inefficiently mapped to network hardware operations. We conclude our findings with four categories of communication needs: asynchrony, concurrency, multithreading, and progress.

The second part of this work is to fix the problem. Existing works have proposed a few techniques to fix or alleviate some of the inefficiencies under different contexts, but we find they usually have limited maturity and availability and thus cannot be directly applied to HPX. In this work, we overcome these limitations by directly supporting the HPX communication abstraction from the lowest public network interface layer (libibverbs for Infiniband and libfabric for Slingshot-11) with the help of an experimental communication library, the Lightweight Communication Interface (LCI)~~\cite{lci17}. We apply techniques including one-sided communication, queue-based completion notification mechanisms, explicit progress engine invocation, avoidance of coarse-grained locks, and replication of communication resources. The resulting LCI parcelport achieves up to 50x better performance in microbenchmarks and 2x in a real-world application than the existing MPI parcelport. 

We further designed LCI parcelport variants to study each technique's performance impact. We find that different techniques have different performance impacts, and combinations of different techniques can also have complicated interactions. For example, the removal of coarse-grained locks itself is not enough to fix the application performance, and we also need to invoke the progress engine at a sufficient frequency; the queue-based completion mechanism is only beneficial when we use state-of-the-art multi-producer-multi-consumer queues; the benefits of one-sided communication primitives are manifested with a lock-free runtime but a well-implemented send/receive can close the gap. We conclude the paper by discussing existing communication libraries and AMT runtimes and how the lessons learned with HPX and LCI can be applied to them.

This project represents an important piece of the larger LCI project, which aims to develop a better communication library for more dynamic and irregular applications. 
This requires a focus shift from raw communication performance (bandwidth and latency) to the mechanisms used by the applications to interact with the communication library.
The design of the LCI parcelport is not constrained by any legacy LCI constructs as we directly modify and augment the LCI interface/runtime whenever needed. This allows for a direct map of AMT communications to native network functionalities. \emph{Therefore, we believe the performance results presented here are not specific to LCI, but reveal fundamental properties of the corresponding communication library designs directly based on libibverbs and libfabrics.} Conceptually, our work can be viewed as equivalent to directly implementing the HPX communication abstraction from the lowest network interface level, while leveraging the LCI infrastructure -- such as low-level network wrappers and bootstrapping -- for convenience.

In summary, this paper makes the following contributions:
\begin{itemize}
\item We articulate the complete picture of the communication stack of a real-world AMT system, HPX, and establish a communication abstraction for underlying communication libraries to support.
\item We identify four categories of inefficiencies in the existing HPX communication layer implementation (the MPI parcelport) due to interface and/or implementation limitations of existing communication libraries.
\item We create a new HPX parcelport (the LCI parcelport) that fixes all the inefficiencies through techniques that map AMT communications more directly to native network layer functionalities
with the help of an experimental communication library, LCI.
\item We demonstrate the performance and scalability of the resulting LCI parcelport, achieving up to 50x message rate improvement, 6x latency reduction, and 2x real-world application speedup compared with the existing HPX+MPI communication stack.
\item We quantify the performance impact of each technique through systematic evaluations of LCI parcelport variants implementing corresponding design options and discuss suggestions for existing communication libraries.
\end{itemize}

\section{The HPX Communication Abstraction}
\label{sec:comm_abs}

We first present an overview of AMTs' communication stack (Section~\ref{sec:comm_stack_overview}) and then describe the functionality covered in the HPX upper communication layer (Section~\ref{sec:upper_layer}). After that, we detail the parcelport communication abstraction that underlying communication libraries need to implement in order to support HPX communications (Section~\ref{sec:parcelport_comm_abs}).

\subsection{Communication Stack Overview}
\label{sec:comm_stack_overview}
\begin{figure}[htbp]
\centering
\includegraphics[width=0.7\linewidth]{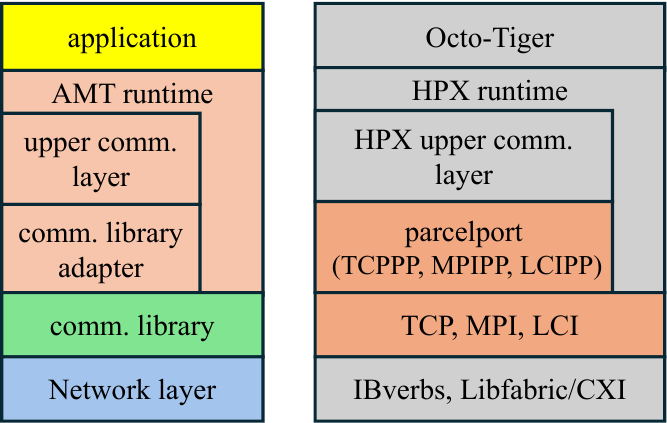}
\caption{AMT Communication Stack}
\label{fig:stack}
\end{figure}

Figure~\ref{fig:stack} illustrates a typical software stack of an AMT. 
The application code sits atop the AMT runtime system. This system includes an upper-level communication layer and an adaptation wrapper that maps the AMT communication abstraction atop a specific communication library. The communication library is implemented on top of a native network layer. We indicate on the right the specific instantiation that we used in our study: \emph{Octo-Tiger}~\cite{marcello2021octo} is an astrophysics application simulating the evolution of binary star systems. Octo-Tiger is implemented atop \emph{HPX}~\cite{Kaiser2020HPX}. HPX extends C++ with constructs that support a global address space (called AGAS~\cite{Amini2019AssessingTP}) and task parallelism. \emph{Parcelport} is the HPX adaptation layer that implements the HPX communication abstraction with a specific communication library. We consider two such: MPI and LCI~\cite{lci17}. We tested on platforms with Infiniband (using Libibverbs)~\cite{Infiniband} or Slingshot-11 (using Libfabric)~\cite{libfabric} interconnect. When running on Infiniband, MPI libraries typically do not directly interact with Libibverbs but instead use an additional portability layer, such as UCX~\cite{shamis2015ucx} or Libfabric~\cite{libfabric}.
On the contrary, LCI is directly built on top of Libibverbs for a broader design space and more explicit network control. Many overheads mentioned in Section~\ref{sec:parcelport_and_commlibs} (such as tag matching and coarse-grained resource locks) also come from this portability layer besides MPI, but we will not always distinguish the portability layer (e.g., UCX, Libfabric) from the communication library layer (e.g. MPI) for simplicity.
This paper focuses on the two layers highlighted in the middle where the interaction between the HPX communication code and the external communication library happens. 
We do not focus on the native network layer, as it only provides basic services and is highly vendor-proprietary, or the upper-level HPX communication code, as it is mostly trivial and implements fundamental functionalities such as serializing/deserializing and addressing.  

\subsection{The HPX Upper Communication Layer}
\label{sec:upper_layer}

\begin{lstlisting}[language=C, caption={The HPX application interface.}, label={lst:app_interface}]
async(locality, func, args...) -> future
async(global_obj, method, args...) -> future
\end{lstlisting}

\subsubsection{Application Interface}

An HPX application comprises a collection of processes referred to by HPX as \emph{localities}. HPX applications can register functions or class methods as \emph{actions}. In the latter case, the corresponding objects are called \emph{global objects}. 
\emph{Actions} can be invoked asynchronously as \emph{tasks} on arbitrary \emph{localities} as shown in Listing~\ref{lst:app_interface}. The first \emph{async} function invokes a function (\emph{func}) with provided arguments (\emph{args}) on the specified process (\emph{locality}). The second \emph{async} function invokes a method (\emph{method}) of a global object (\emph{global\_obj}) with provided arguments (\emph{args}) on the process hosting the global object. The returned \emph{future} handlers can be composed into arbitrary task dependency graphs.
Communication happens during inter-process task spawning and synchronization, essentially in the form of Remote Procedure Calls (RPCs).

\subsubsection{Runtime Design}

Internally, HPX maintains a collection of \emph{worker threads}, implemented as C++ threads, to execute those invoked actions (\emph{tasks}). By default, each worker thread is pinned to a distinct CPU core. 
A communication request can be spawned from any worker thread and served by any worker thread on the target locality.

The HPX upper communication layer implements three communi\-cation-related services: global object addressing, argument serialization, and parcel aggregation. We will discuss the latter two on the sender side. The receiver side is symmetric.

\paragraph{Argument Serialization. } The metadata of the action invoked and the arguments are serialized into an HPX-internal data structure called \emph{parcel}, which consists of one or more consecutive buffers (\emph{chunks}) as follows:
\begin{itemize}
    \item  A \emph{data chunk} containing the metadata of the action and all the small arguments.
    \item Optionally, multiple \emph{zero-copy chunks}, each containing a large argument of the action.
    \item A \emph{transmission chunk} containing the index and length of the serialized arguments. It is only needed when there is at least one zero-copy chunk. 
\end{itemize}
An internal parameter, the HPX zero-copy serialization threshold, dictates whether an argument is small or large.

\paragraph{Parcel Aggregation. } The HPX runtime opportunistically aggregates parcels sharing the same destination to reduce communication frequency. The HPX runtime maintains one parcel queue per destination locality. To send a parcel, the upper layer first enqueues the parcel into the corresponding parcel queue. Next, it dequeues all the parcels in that parcel queue, aggregates them into one parcel, and passes it to the parcelport layer. This creates opportunities for message aggregation when multiple threads push parcels to the same parcel queue simultaneously.

\subsection{The Parcelport Communication Abstraction}
\label{sec:parcelport_comm_abs}

The parcelport layer transfers the serialized and potentially aggregated parcels to the target locality. Each communication library requires its own parcelport. To simplify the description, we merge the \emph{data chunk} and the \emph{transmission chunk} into a single \emph{nonzero-copy chunk}. Listing~\ref{lst:parcelport_interface} shows the communication abstraction that the parcelport layer needs to support.

\begin{lstlisting}[language=C, caption={The HPX communication abstraction.}, label={lst:parcelport_interface}]
struct Chunk { char address[]; size_t size; }
struct Parcel { 
  Chunk nzc_chunk;
  Chunk zc_chunks[]; size_t num; 
}
send(locality, parcel, cb) -> void
background_work() -> bool
// Provided by the upper layer
allocate_zc_chunks(nzc_chunk) -> parcel
handle_parcel(parcel) -> void
\end{lstlisting}

A \emph{parcel} consists of a \emph{nonzero-copy chunk} and an arbitrary number of \emph{zero-copy chunks}. The later chunks can be of arbitrary size. The upper layer must allocate the receive buffers for the zero-copy chunks through the \emph{allocate\_zc\_chunks} as many C++ STD data structures cannot be constructed from existing memory buffers. The \emph{allocate\_zc\_chunks} takes the \emph{nonzero-copy chunk} as an input argument as it contains the size information of zero-copy chunks.

The parcelport implements two functions: \emph{send} and \emph{background\_work}. Any HPX worker thread can invoke \emph{send} to send a \emph{parcel}. The callback function \emph{cb} is invoked after the send completes. A typical callback releases the temporary buffers associated with the parcel. The workers repeatedly call the \emph{background\_work} function when idle to check for new incoming parcels and make progress on pending parcel transfers. The return value of \emph{background\_work} indicates whether any progress is made, serving as a hint for the upper layer. When the parcelport layer receives a parcel, it will call the \emph{handle\_parcel} function to deliver the parcel to the upper layer.
\section{Parcelport and Communication Libraries}
\label{sec:parcelport_and_commlibs}

In this section, we look at the design of the existing MPI parcelport and the new LCI parcelport. We will take a holistic view across native network interfaces, communication libraries, and HPX parcelports to analyze the inefficiencies inside the MPI parcelport and discuss how we fix them in the LCI parcelport. We will conclude with four categories of communication needs of AMTs for communication libraries. This section analyzes these communication needs from basic principles, and section~\ref{sec:factor_study} evaluates them experimentally. A full description of LCI's design is out of the scope of this paper, but we will describe how the corresponding LCI features map to native network layer behaviors as needed.

Section~\ref{sec:native_network_layer} first briefly overviews the native network layer to establish the necessary background. Section~\ref{sec:parcelport_common_overview} then presents the common high-level overview of the two parcelports. Lastly, Section~\ref{sec:parcelport_needs} examines each design difference between the two parcelports in detail and articulates the communication needs.

\subsection{The Native Network Layer}
\label{sec:native_network_layer}

To better understand what the communication libraries do, we first present briefly the native network interface. We focus on \emph{Libibverbs} in this section as Infiniband is a widely used interconnect and is well-understood by the research community.

\emph{Libibverbs} processes use the following data structures to communicate with each other: a set of queue pairs, at least one completion queue, and one or more optional shared receive queues. A \emph{queue pair} consists of a \emph{send queue} and a \emph{receive queue}. It represents a logical communication channel between two processes. Processes issue communication operations by posting message descriptors to a corresponding send or receive queue. 
Commonly, \emph{Libibverbs} users create \emph{shared receive queues} to serve as a common receive queue for all queue pairs. 
All communication operations are asynchronous. Every send and receive queue is associated with a \emph{completion queue}. The hardware notifies users of completed operations by pushing completion descriptors to completion queues, and users need to poll the completion queue with sufficient frequency to avoid overflow.

\emph{Libibverbs} supports two-sided send/receive
and one-sided RDMA write/read of remote buffers. The RDMA write can carry an optional 4-byte signal for remote completion notification. Unexpected sends will trigger RNR (Receive Not Ready) errors, which are disastrous to application performance~\cite{kong2023UnderstandingRDMAMicroarchitecture}. All communication buffers need to be registered to the hardware. For RDMA operations, the caller needs to specify the remote keys of remote buffers.

While \emph{libfabric} offers a more versatile API and (when running on Slingshot-11) the native network layer (the HPE Cassini interface) is not publicly documented, we suspect it operates in a similar manner: there are some queues for posting communication requests and completion polling; basic primitives includes send/receive/write/read; unexpected sends are not favorable; RDMA buffers need to be registered, etc.

\subsection{Common High-level Overview}
\label{sec:parcelport_common_overview}

The communication abstraction described in Section~\ref{sec:parcelport_comm_abs} defines the core interface needed to support HPX. In practice, both parcelports generate an additional \emph{header} for each parcel. The header contains metadata about the parcel, such as the tag used by the follow-up sends/receives and the nonzero-copy chunk size. Eventually, a parcel is communicated by both parcelports with the following sequence of messages: one \emph{header message}, one \emph{nonzero-copy chunk message}, and zero or more \emph{zero-copy chunk messages}. We will call the latter two types of messages the \emph{follow-up} messages. If the nonzero-copy chunk messages are small enough, they will be piggybacked onto the header message.

Distinct parcels can be transferred concurrently, but chunks of the same parcel are transferred sequentially: The sender starts to send a new chunk only after the sending of the prior chunk of the same parcel has been completed; likewise, the receiver starts to receive a new chunk only after the receiving of the prior chunk has been completed. All the parcelport communication operations are nonblocking for communication overlap. Therefore, parcelports need to periodically check the completion status of the posted nonblocking operations and, if completed, take the subsequent actions, such as posting new communication operations or notifying the upper layer.

\subsection{Design Details and Analysis}
\label{sec:parcelport_needs}
\subsubsection{Header Message Transfer}
\label{sec:parcelport_asynchrony}

In HPX, task signaling and dependency fulfillment are one-sided. When mapped to the parcelport level, only the senders are aware of sending a parcel, while the receivers remain unaware until the parcel arrives.

Formally, the header message transfer of a parcel has the following three properties. (a) \emph{size bounded}: the message size has a fixed upper bound. (b) \emph{unexpected}: the receiver does not know whether or when it will receive such messages. (c) \emph{location agnostic}: the receiver does not care about the exact location of the receive buffer.

The current MPI parcelport receives the header messages by pre-posting an \emph{MPI\_Irecv} with \emph{MPI\_ANY\_SOURCE} on each process and using the \emph{background\_work} (Listing~\ref{lst:parcelport_interface}) to poll the resulting request periodically. Once the request is detected as ready, the runtime will pre-post another any-source receive and then process the incoming header message.

To avoid the RNR error in the native network layer, communication libraries (specifically, their progress engine) typically pre-post many native network receives and periodically poll network completion queues to check for completed ones. Therefore, the low-level network interface inside communication libraries is already a perfect match for header message transfer: the runtime just needs to deliver the received messages to the users through some completion mechanism. However, the restriction of the MPI interface forces users to pre-post \emph{MPI\_Irecv} for such messages. This introduces unnecessary overhead such as tag matching. In addition, the concurrency of processing the received messages is restricted. Since only one \emph{MPI\_Irecv} is pre-posted at any given time, only one thread can proceed along the code path inside \emph{MPI\_Irecv} from tag matching to completion signaling. It becomes a sequential bottleneck.

The LCI parcelport fixes these inefficiencies by using the LCI one-sided \emph{dynamic put} primitive. It is similar to the active message primitive but uses a completion queue, instead of a callback, as the remote completion mechanism (Section~\ref{sec:parcelport_concurrency} discusses the reason). Once a corresponding completion descriptor is popped from the native network completion queue, the runtime directly hands the internal receive buffer to the users through an atomic-based LCI completion queue, reducing the overhead to the bare minimum.

The follow-up messages are of arbitrary size and are always expected. The upper communication layer needs to allocate the receive buffers of the \emph{zero-copy chunk messages}. As a result, both parcelports use the two-sided nonblocking send/receive communication primitives. 

\req{\textbf{Asynchrony Support:}
AMT control messages are unexpected by the receivers. They are best supported by a one-sided communication primitive that can directly deliver the message to the receipt. The traditional tag-matching send-receive is superfluous.
}

\subsubsection{Managing many pending operations}
\label{sec:parcelport_concurrency}

HPX executes many tasks simultaneously, resulting in many pending communication operations. The parcelport is event-driven: it does not care about the ordering in which these pending operations are completed; it just needs to perform certain actions whenever an operation is completed. 

The only completion notification mechanism available in the MPI standard today is to query the per-operation \emph{MPI\_Request} object. In HPX, the MPI parcelport stores the pending requests in two shared request pools, one for sends and the other for follow-up receives, implemented with C++ \emph{deque} and HPX try-lock. Each call to the \emph{background\_work} function will check one request in the pools using \emph{MPI\_Test} in round-robin order. It does not use \emph{MPI\_Testsome} for two reasons: (a) managing the input request array in the critical section would be expensive, (b) MPI vendors typically just implement \emph{MPI\_Testsome} as a for loop of \emph{MPI\_Test} traversing the request array.

However, lower-level communication layers typically have a direct mechanism to quickly poll for an arbitrary completed operation, for example, the completion queue in Libibverbs/Libfabric or the completion handler in UCX. It is inefficient for MPI clients to loop over multiple requests to check the completion status of individual operations. The extra synchronization efforts involved in the multithreaded case can further reduce performance.

We use a queue-based completion mechanism in the LCI parcelport to deliver completed operations to the parcelport layer. This removes the locking overhead and reduces the waiting time for a completed operation to be handled. By default, the LCI completion queues use LCRQ~\cite{Morrison2013lcrq}, the state-of-the-art Multi-Producer-Multi-Consumer (MPMC) lock-free linearizable queue supporting dynamic resizing. 

LCI also offers a callback-based completion mechanism besides the queue-based one, as those in UCX. However, the handling of received parcels (\emph{handle\_parcel} in Listing~\ref{lst:parcelport_interface}) may invoke arbitrary user-defined tasks, which can potentially be blocked and de-scheduled by the HPX scheduler. The callback handlers are invoked within the progress engine of the communication runtime. Executing such procedures in this context could disrupt the normal operation of the progress engine, potentially leading to deadlock~\cite{bonachea2002gasnet}. Therefore, we use the queue-based mechanism to decouple the handling of HPX messages from the communication runtime context.

\req{\textbf{Concurrency Support:} 
AMTs need an efficient mechanism to poll for a completed operation among many pending ones. This need stems from the common use of task oversubscription in AMTs to increase computation/communication overlap and reduce idle time. A queue-based completion mechanism is recommended while the callback-based one is not directly applicable due to the dynamicity of the task runtime.}

\subsubsection{Multithreading Efficiency}
\label{sec:parcelport_threading}

In HPX, all worker threads can call the \emph{send} and \emph{background\_work} functions in Listing~\ref{lst:parcelport_interface}, so the communication invocations are heavily multithreaded.
 
From the native network layer perspective, multiple operations can be concurrently executed. Native network resources, such as \emph{libibverbs} queues,
typically use distinct locks to ensure thread safety. Moreover, each logical resource can typically be mapped to an individual set of hardware resources (e.g., uUARs for Infiniband~\cite{zambre2018ScalableCommunicationEndpoints}), supporting nonconflicting concurrent communications directly at the hardware level.

For the communication library layer, this contention isolation means that threads performing different actions (\emph{send}, \emph{receive}, \emph{progress}) do not necessarily need to compete with each other for low-level resources. In most cases, a \emph{send} only needs to access a send queue; a \emph{receive} only needs to access the tag-matching data structure; and a \emph{progress} (used to perform background works) only needs to access the completion queues and receive queues (to pre-post network receives).
However, existing communication libraries commonly wrap a complete set of these resources in a single data structure and protect it with a coarse-grained blocking lock. This destroys the independence between threads performing non-conflicting actions. 

LCI and the LCI parcelport fix this inefficiency in two ways. First, LCI removes the coarse-grained locks that wrap the network resources. It directly exposes the communication concurrency of the native network layer to its clients. Second, LCI explicitly represents the complete set of the network resources in the LCI \emph{device} data structure. The LCI parcelport can utilize multiple LCI devices to reduce the contention between threads simultaneously calling \emph{progress} or calling \emph{send} with the same target rank. 

We present here how the LCI parcelport utilizes multiple LCI devices. All the messages of the same parcel are sent using the same device. We use a static mapping from worker threads to devices to determine the device a worker thread uses to send parcels and make progress. We still use a shared LCI completion queue across all devices to reduce load imbalance, so a worker thread may still need to access other devices to post receives and follow-up sends to the appropriate device. The header message carries the device index to be used for the receives.

\req{\textbf{Multithreading Support:} AMTs need efficient support for multithreaded communication due to the fact that AMTs are generally multithreaded and many threads can communicate simultaneously. Two approaches can help: expose hardware-level concurrency to improve the threading efficiency within the device and allocate more resources to alleviate the contention further.}

\subsubsection{Driving the progress engine}
\label{sec:parcelport_progress}

Progress is an important concept in communication libraries~\cite{Hoefler2008mpi_progress_thread}. The MPI specification does not define an explicit progress function. Therefore, MPI progress engine invocation happens as a side-effect of certain MPI calls, and users do not have explicit control over when and how often the MPI progress engine is invoked. Current MPICH and OpenMPI implementation only poll the progress engine during calls to \emph{MPI\_Test}. 
Moreover, multiple threads cannot simultaneously test an MPI request even with \emph{MPI\_THREAD\_MULTIPLE}~\cite{mpi41_shared_request}. Hence, the MPI parcelport has to wrap a lock around every \emph{MPI\_Test}. "Gluing" together two distinct functions and having them protected by the same lock is a source of inefficiency that limits progress frequency.

These limitations are purely in the communication library layer and have nothing to do with the hardware or low-level communication interface. We remove these limitations in LCI and offer users an explicit \emph{progress} function. The LCI parcelport invokes this \emph{progress} function every time the HPX \emph{background\_work} function is invoked. 

In addition, the LCI parcelport also has the option to allocate dedicated CPU cores for communication progress. This could ensure stable progress and reduce the thread contention on the progress engine, but we have not found sufficient evidence through experiments to justify this design option. We omit the experiment results in Section~\ref{sec:factor_study} due to the page limit.

\req{\textbf{Progress Support:} An explicit progress function gives AMTs more control over when and how frequently to make progress on the communication runtime. It avoids the slowing down of pending communication due to insufficient progress.}

\section{Performance Evaluation}
\label{sec:evaluate}
In this section, we use a parameterized HPX communication micro-benchmark and a real-world application, Octo-Tiger, to showcase the overall performance of the LCI parcelport. We will also briefly talk about existing publications~\cite{strack2024hpx_fft,daiss2024octotiger} that use the LCI parcelport with more applications and clusters. Section~\ref{sec:factor_study} then quantifies the performance contribution of each technique and its alternatives.

\begin{table}[htbp]
\caption{Platform Configuration.}
\label{table:platform_config}
\begin{center}
\small
\begin{tabular}{llll}
\toprule
Platform & SDSC Expanse & TACC Frontera & NCSA Delta \\
\midrule
 CPU & AMD EPYC 7742 & Intel Cascade Lake & AMD EPYC 7763 \\
  & (2 sockets/node) & (2 sockets/node) & (2 sockets/node)\\
  & (128 cores/node) & (56 cores/node) & (128 cores/node)\\
 NIC &  Mellanox CX-6 &  Mellanox CX-6 & Cassini\\
 Network & HDR InfiniBand  & HDR InfiniBand & Slingshot-11 \\
 & (2x50Gbps) & (2x50Gbps) & (200Gbps) \\
 Software & OpenMPI 4.1.5 & Intel MPI 21.9.0 & Cray MPICH 8.1.27 \\
 & UCX 1.14.0 & Libfabric 1.13 & Libfabric 1.15.2.0 \\
 & Libibverbs 43.0 & Libibverbs 43.0 & \\
\bottomrule
\end{tabular}
\end{center}
\end{table}

\subsection{Experimental Setup}

We run all experiments on SDSC Expanse, TACC Frontera, and NCSA Delta. Their configurations are shown in Table~\ref{table:platform_config}. Most experiments show similar trends across three platforms. Due to page limits, we focus on the results from SDSC Expanse and talk about results on other platforms when needed. All experiments are conducted at least five times except for some large-scale runs. The figures show the average and standard deviation. 

\subsection{Experiment Result}
\label{sec:evaluate_mpi}

HPX parcel aggregation exhibits a trade-off between lower parcelport stress due to reduced message numbers and additional thread contention on the parcel queues. In practice, parcel aggregation can sometimes help the MPI parcelport but never helps the LCI parcelport. Therefore, we show results for two MPI parcelport configurations: one with the parcel aggregation disabled (\emph{mpi}) and the other with the parcel aggregation enabled (\emph{mpi\_a}). The LCI parcelport configuration evaluated here always has parcel aggregation disabled and uses two devices. 

\subsubsection{HPX Micro-benchmarks}

\begin{figure}[htbp]
  \centering
  \begin{subfigure}[b]{\linewidth}
    \includegraphics[width=\textwidth]{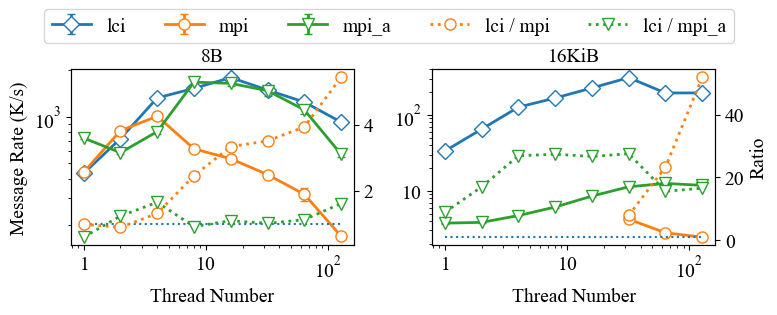}
    \caption{Message Rate Micro-benchmark from 1 to 128 threads.}
    \label{fig:microbenchmarks-msg_rate}
  \end{subfigure}
  \begin{subfigure}[b]{\linewidth}
    \includegraphics[width=\textwidth]{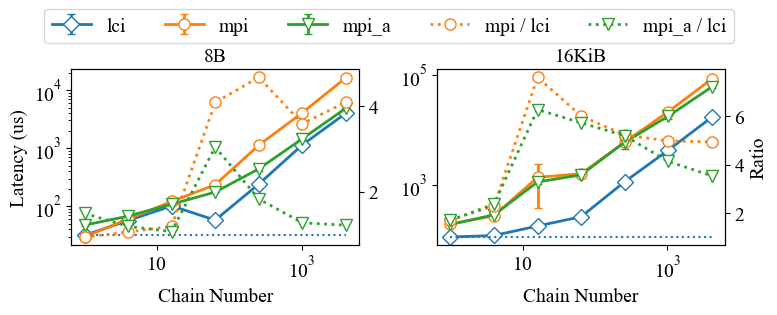}
    \caption{Latency Micro-benchmark from 1 to 4096 chains.}
    \label{fig:microbenchmarks-latency}
  \end{subfigure}
  \caption{Microbenchmark results of 8B/16KiB messages on Expanse: the solid lines/left axis show the absolute message rates/Latency; the dotted lines/right axis show the relative speedup. \emph{mpi\_a} means MPI with aggregation.}
\end{figure}

Task systems' messages tend to be small and typically do not saturate the network bandwidth. Therefore, we design our micro-benchmarks to stress the underlying network stack with relatively small messages (up to 16 KiB) and focus on latency and message rate.

We use a parameterized task graph for the micro-benchmarks. The task graph consists of a fixed number (\emph{nchains}) of chains, each with a fixed number (\emph{nsteps} + 1) of tasks. The program runs on two HPX processes, with tasks of a chain alternating between the two processes. Task arguments have fixed size (\emph{msg\_size}). The two processes run on two separate nodes, and the number of HPX threads by default equals the number of available cores. 
For the message rate micro-benchmark 
, we set \emph{nchains} to be very large and \emph{nsteps} to one so that we have one process (the sender) sending a flood of messages to the other (the receiver). 
For the latency micro-benchmark 
, we set \emph{nsteps} to be very large and \emph{nchains} to be relatively small so that we have a multi-message ping-pong-like program.

Figure~\ref{fig:microbenchmarks-msg_rate} shows the message rates achieved by different thread numbers.
Figure~\ref{fig:microbenchmarks-latency} shows the latency for various numbers of concurrent chains, evaluating the parcelports' ability to overlap concurrent communications. Both micro-benchmarks use two message sizes: 8B and 16 KiB. With 8B messages, each parcel is transferred through one message: the header message. With 16KiB messages, each parcel is transferred through one header message and one follow-up message. We only show Expanse results here but results on Frontera and Delta are qualitatively similar.
We make the following observations:

\textbf{The LCI parcelport outperforms both MPI parcelport variants, usually by a significant margin.} Compared to the best variant of the MPI parcelport, \emph{lci} achieves a message rate improvement and latency reduction of up to 3x for short messages and up to 20x for long messages. 

\textbf{Parcel aggregation yields mixed results for the MPI parcelport.} The variant with aggregation enabled generally shows better performance for micro-benchmarks, with an improvement of around 3x compared to the variant with no aggregation. However, aggregation sometimes hurts the performance of large messages, resulting in an astonishing 50x slow down of \emph{mpi\_a} compared to \emph{lci} large message rate. It is expected that aggregation helps more with small parcels than large parcels, as it cannot combine the zero-copy chunks. 

\textbf{A single thread is not enough to achieve maximum communication performance.} 
Communication funneling is an alternative approach for MPI clients to workaround the inefficiency of \emph{MPI\_THREAD\_MULTIPLE}. As there is no computation in the microbenchmark, we could use the single-threaded performance to roughly estimate what will happen if we funnel all communication to one thread. Comparing the single-threaded and multithreaded message rate in Figure~\ref{fig:microbenchmarks-msg_rate}, we see that this strategy can be preferable with non-aggregated MPI and short-message-dominated workloads, but not with LCI due to its improved multithreading support, where using more threads can improve message rate by 4x.

\subsubsection{Application Benchmark}

We use the Octo-Tiger~\cite{marcello2021octo} application code as a benchmark to understand performance for a realistic workload. Octo-Tiger is an astrophysics application simulating the evolution of binary star systems based on the fast multipole method on adaptive octrees. It is built on top of HPX and extensively uses HPX actions and local control objects to achieve asynchronous execution and computation-communication overlap.
We use the "rotating star" scenarios and set the maximum octree depth to five. Each octree leaf holds an $8 \times 8 \times 8$ mesh. 
We run 5 time steps.
The HPX zero-copy threshold is set to 8KiB. We run two processes per node on Expanse and one process per node on Frontera to maintain a similar thread number per process across clusters. 

\begin{figure}[htbp]
  \centering
  \includegraphics[width=\linewidth]{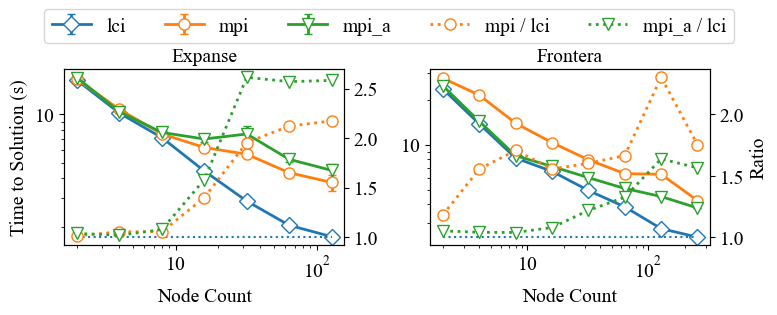}
  \caption{Execution time and performance ratio of Octo-Tiger, strong scaling up to 128/256 nodes on Expanse/Frontera.}
  \label{fig:octotiger}
\end{figure}

Figure~\ref{fig:octotiger} shows the total execution time of Octo-Tiger with different numbers of computation nodes on Expanse and Frontera. 
On both platforms, the LCI parcelport achieves increasing speedups relative to the MPI parcelport as the node count increases. Compared to \emph{mpi}, the MPI parcelport without parcel aggregation, we get up to 2x speedup. Compared to \emph{mpi\_a}, the MPI parcelport with parcel aggregation enabled, we get up to 2.5x speedup. Aggregation helps the MPI parcelport on Frontera but worsens it on Expanse.

In addition, \cite{strack2024hpx_fft} developed a multidimensional Fast Fourier Transform solver on top of HPX and reported 5x speedup with the LCI parcelport over the MPI parcelport and the FFTW baseline.

\subsubsection{Slingshot-11 Result}
\label{sec:eval_ss11}

\begin{figure}[htbp]
    \centering
    \begin{subfigure}{0.49\linewidth}
        \includegraphics[width=\linewidth]{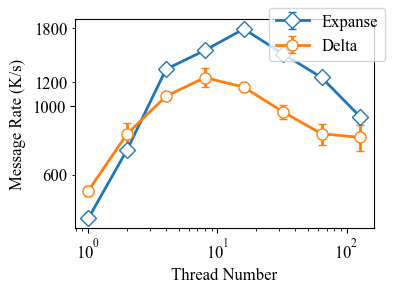}
        \caption{Message rate (8B)}
    \end{subfigure}
    \begin{subfigure}{0.49\linewidth}
        \includegraphics[width=\linewidth]{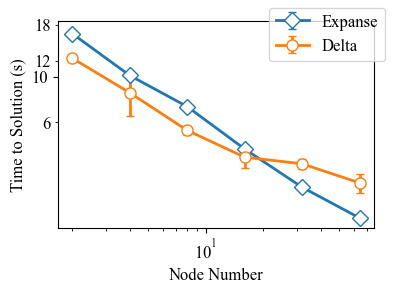}
        \caption{Octo-Tiger Time}
    \end{subfigure}
    \caption{Performance comparison of Expanse (Infiniband) and Delta (Slingshot-11) with the HPX LCI parcelport.}
    \label{fig:ss11}
\end{figure}

When running on Slingshot-11, MPI and LCI are both based on \emph{Libfabric}. Figure~\ref{fig:ss11} shows the message rate (8B) micro-benchmark and Octo-Tiger results comparing results on Expanse and Delta. Compared to Expanse, Delta has similar CPU processors and memory with a theoretical higher-performance interconnect. However, it achieves 30\% lower peak small message rates and 33\% lower application performance (with 64 nodes). We suspect the main performance bottleneck here is the coarse-grained blocking lock inside Libfabric. \emph{Perf} reports that when running Octo-Tiger on Delta with 32 nodes, 85\% execution time is spent on a \emph{pthread\_spin\_lock} inside the Libfabric completion queue polling.

While the Libfabric blocking locks handicap both LCI and MPI, the LCI parcelport still achieves 1.2x-3x Octo-Tiger speedup compared to the MPI parcelports on Delta with 32/64 nodes. \cite{daiss2024octotiger} has conducted larger-scale production-level experiments of Octo-Tiger with the LCI parcelport on Perlmutter with Slingshot-11. They scale the application up to 1700 GPU nodes, where the LCI parcelport outperforms the MPI parcelport with a time-to-solution reduction of 2x. All MPI results reported here use the recommended Cray-MPICH.
\section{Quantifying Contributions}
\label{sec:factor_study}

We now study the performance contribution of each technique. We implement multiple LCI parcelport variants, each differing in one or a few features. 
We use the same micro-benchmarks and application as in the previous section. All experiments shown in this section are run on Expanse. Because of the large number of variants to evaluate, we focus on the following five metrics: the message rates for small messages (8B) and large messages (16KiB) with one thread per core, the latency of small messages (8B) and large messages (16KiB) with 1024 chains, and the total execution time of Octo-Tiger on 32 nodes. 
We take the inverse of the latency and execution time to correlate all presented numbers with performance positively. We further normalize the performance numbers against a base case so that a number larger than one means speedup and a number smaller than one means slowdown. All experiments are performed at least five times, and we present the average and standard deviation. Unless otherwise noted, the base case uses the same configuration as the \emph{lci} presented in Section~\ref{sec:evaluate_mpi}: \emph{dynamic put} for header messages and \emph{send}-\emph{receive} for followup messages; completion queues; all worker threads explicitly calling \emph{LCI\_progress} when idle; coarse-grained-lock-free communication runtime; and two devices.

\subsection{Asynchrony}
Dynamic MPI applications commonly pre-post \emph{MPI\_Irecv} and periodically test them to check for unexpected incoming messages. As Section~\ref{sec:parcelport_asynchrony} analyzes, besides complicating the codebase, there are two potential performance disadvantages: the use of an additional operation and the serialization of the processing of incoming messages. In the LCI parcelport, we use the LCI \emph{dynamic put} primitive to inject messages to target completion queues directly. We investigate here how much real-world performance improvement we can get from this ideal one-sided primitive (dynamic put) compared to the more commonly used two-sided primitive (send/recv).

\begin{figure}[htbp]
    \centering
    \includegraphics[width=0.9\linewidth]{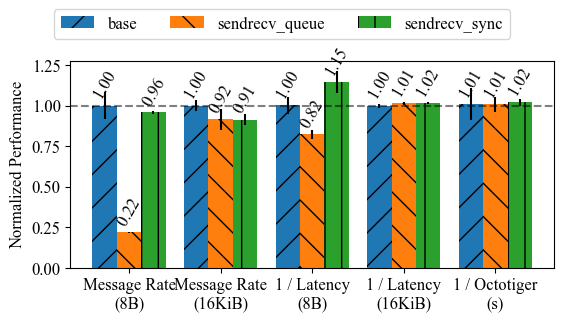}
    \caption{Factor study for asynchrony support features.
    \emph{base} uses the ideal \emph{dynamic put} primitive with a completion queue as the target.
    \emph{sendrecv\_queue} uses \emph{send/recv} and keeps the completion queue. 
    \emph{sendrecv\_sync} replaces the completion queue with a synchronizer.
    }
    \label{fig:factor_study-asynchrony}
\end{figure}

We implement LCI parcelport variants that use LCI send/recv instead of \emph{put dynamic} to transfer the header messages and use a queue-based tag matching scheme similar to that used by MPI (\emph{sendrecv\_base}). We also replace the target completion queues with an LCI synchronizer, similar to an MPI request, to reduce the signaling overhead (\emph{sendrecv\_sync}). In effect, we specialize the completion queue to the case where it will never contain more than one entry.

Figure~\ref{fig:factor_study-asynchrony} shows the experiment results. Replacing the \emph{dynamic put} with \emph{send/recv} worsens the performance of small messages (78\% message rate decrease) but has no significant impact at the application level. However, the performance gap for small messages is closed after replacing the target completion object from a completion queue to a synchronizer. The send/recv synchronizer-based approach even achieves slightly better latency (15\%) than the put-queue-based approach. This is due to three facts: (a) pushing into a completion queue is more expensive than signaling a synchronizer (i.e., a queue with only one slot) (b) compared with \emph{put dynamic}, pre-posting only one receive limits the concurrency of incoming message processing so that the latency of the signaling mechanism becomes more critical (c) other receive-related overheads, such as tag matching and memory copy, do not significantly affect AMT performance.

\obs{The ideal one-sided primitive shows advantages in AMTs' microbenchmarks in certain cases, but these advantages can be insignificant given an efficient receive implementation. Preposting receives limit the concurrency of incoming message processing so that performance is more sensitive to the efficiency of the receive implementation.}

\subsection{Concurrency}
\label{sec:factor_study-concurrency}
As Section~\ref{sec:parcelport_concurrency} shows, a request-based completion mechanism becomes cumbersome when there are many pending operations. In addition, directly applying a callback-based completion mechanism is not practical due to complicated runtime logic. Therefore, we choose to use completion queues to save the overhead related to request pool management. We investigate here whether there are any visible performance impacts of using a queue-based completion mechanism over the request pool approach at the AMT user level.

\begin{figure}[htbp]
    \centering
    \includegraphics[width=\linewidth]{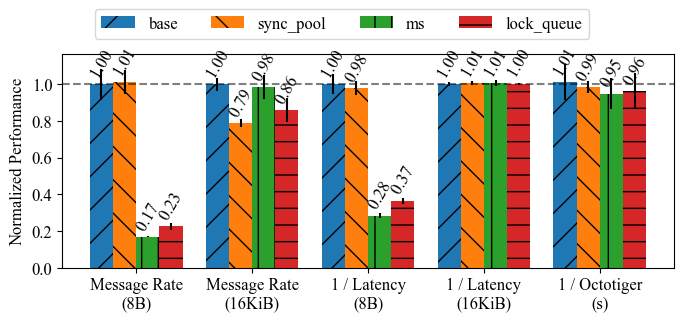}
    \caption{Factor study for concurrency support features.
    \emph{base} uses the default LCRQ as its completion queue.
    \emph{sync} uses a synchronizer pool. 
    \emph{lock\_queue} uses a lock-based completion queue. 
    \emph{ms} uses a CAS-based link-list queue (Michael-Scott Queue~\cite{michael1996msqueue}). 
    }
    \label{fig:factor_study-concurrency}
\end{figure}

We implement LCI parcelport variants that use LCI synchronizers instead of completion queues for all operations other than the pre-posted receives for the header messages (\emph{sync}). The synchronizer pool is implemented similarly to the request pool in the HPX MPI parcelport. By default, LCI uses a highly optimized FAA-based array queue (LCRQ~\cite{Morrison2013lcrq}) as its completion queue. We replace it with lock-based queues (\emph{queue\_lock}) and CAS-based link-list queues (Michael-Scott Queue~\cite{michael1996msqueue}, \emph{queue\_ms}) to examine the impact of the queue implementation.

Figure~\ref{fig:factor_study-concurrency} shows the experiment results. Replacing the completion queues with synchronizer pools causes around 20\% drop in the maximum message rate of large parcels. This shows the performance advantages of the queue-based completion mechanism. Small parcel only consists of one header message (which still uses completion queue as the target for \emph{dynamic put}) so \emph{sync} does not impact their performance. The efficiency of the completion queue implementation also matters: an inefficient queue, such as the lock-based implementation and even the CAS-based Michael-Scott queue, is not enough to achieve maximum performance. 

\obs{Having an efficient completion mechanism to poll for many pending operations is beneficial at the AMT level. A queue-based mechanism shows real-world improvement over the classical request pools. A highly optimized queue implementation is needed to achieve these benefits.}

\subsection{Multithreading and Progress}
\label{sec:factor_study-threading}

Many communication libraries allocate a single set of communication resources per process and use coarse-grained blocking locks to ensure thread safety. This can cause severe thread contention at both the communication library layer and lower layers. LCI and the LCI parcelport use two approaches to fix this problem: remove the coarse-grained lock at the communication library layer and allocate multiple independent communication resources. A less intrusive change is to replace the blocking locks inside the progress engine with try locks.\footnote{Replacing the blocking lock inside send/recv is more difficult as MPI Isend/Irecv does not define a return value to reflect a failed try lock.} 
This pays off since the HPX runtime usually has alternative work if it cannot advance communication.

The MPI standard does not define an explicit way to invoke its progress engine; it is usually invoked implicitly when an \emph{MPI\_Test} call returns \emph{false}. An explicit progress function might benefit performance by providing more control on progress invocations and reducing the overhead of completion tests. 
Here, we investigate the performance impacts of multithreading and progress support techniques. We discuss them together as we find they can have synergetic effects.

\begin{figure}[htbp]
    \centering
    \includegraphics[width=\linewidth]{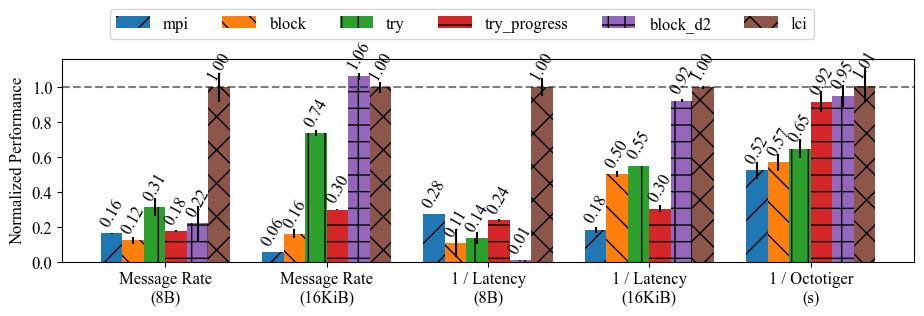}
    \caption{Factor study for the multithreading and progress support features.
    \emph{mpi} uses the MPI parcelport. 
    \emph{block} uses one device with coarse-grained blocking locks. 
    \emph{try} uses one device with coarse-grained try locks. 
    \emph{try\_progress} uses one device with coarse-grained try locks and explicit progressing. 
    \emph{block\_d2} uses two devices with coarse-grained blocking locks. 
    \emph{lci} uses two devices with fine-grained locks.
    }
    \label{fig:factor_study-threading0}
\end{figure}

We first study the impact of try locks and resource replication. We implement LCI configurations that wrap an LCI device with a coarse-grained lock (either blocking or try lock). As we have seen in Sec.\ref{sec:factor_study-concurrency}, completion queues do not work well with locks, so we use send/recv and synchronizers in this experiment. In addition, we find that progress frequency plays an important role when there are coarse-grained locks, so we also add LCI parcelport variants that only invoke progress implicitly when synchronizer testing returns "not ready". In summary, We test the following configurations:
\begin{itemize}
    \item \emph{mpi}: the MPI parcelport.
    \item \emph{block}: the LCI parcelport variant mimics the MPI parcelport in various aspects including send/receive for header messages, synchronizer for completion polling, progress engine invocation only when synchronizer testing fails, synchronizer testing protected by HPX try locks, one device, and a coarse-grained blocking lock wrapping the device.
    \item \emph{try}: similar to \emph{block} but using \emph{try\_lock} for the progress engine invocation.
    \item \emph{try\_progress}: similar to \emph{try} but using explicit progress function to improve progressing frequency.
    \item \emph{block\_d2}: similar to \emph{block} but using two devices.
    \item \emph{lci}: the full-fledged LCI parcelport.
\end{itemize}

Figure~\ref{fig:factor_study-threading0} shows the experiment results. For Octo-Tiger, we can see that the \emph{block} variant achieves similar performance to the MPI parcelport and \emph{try\_progress}/\emph{block\_d2} achieves similar performance to \emph{lci}. This shows us two complementary ways to improve application performance. First, we could switch to \emph{try\_lock} inside the progress engine and increase the progress frequency with an explicit progress engine  (\emph{try\_progress}). Both factors are important: only using \emph{try\_lock} (\emph{try}) results in moderate improvement. Second, we could replicate the communication devices to reduce the thread contention on the coarse-grained lock. Two devices are enough to close the performance gap at the application level. However, for microbenchmarks, even though the above approaches still improve performance, there are still large performance gaps between them and \emph{lci}.

We also find mutual effects between blocking locks and explicit progress. We tested an LCI parcelport variant \emph{progress}: It is similar to \emph{block} (so still with blocking locks) but uses an explicit progress function to improve progress frequency. This results in devastating performance (not shown here due to job timeout) as all threads are waiting in the progress function trying to acquire the lock. \emph{Invoking progress more frequently pays off only if the progress function does not use a coarse blocking lock.} The MPI parcelport, on the other side, does not fall into this trap due to two facts: (a) MPI will only invoke the progress engine when testing the request object (b) MPI disallows concurrency testing of shared request~\cite{mpi41_shared_request} so HPX wraps a try lock around the request testing. It is fun to learn that two seemingly inefficient design choices in MPI (no explicit progress and no concurrent request testing) lead to a positive effect in combination.

\begin{figure}[htbp]
    \centering
    \begin{subfigure}{0.49\linewidth}
        \includegraphics[width=\linewidth]{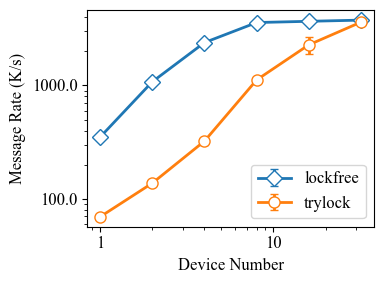}
        \caption{Message Rate (8B).}
    \end{subfigure}
    \begin{subfigure}{0.49\linewidth}
        \includegraphics[width=\linewidth]{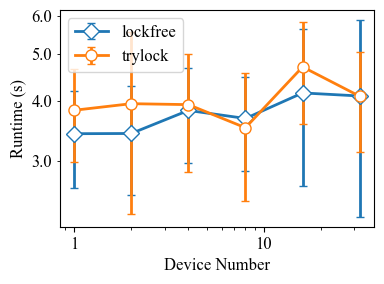}
        \caption{Octo-Tiger time.}
    \end{subfigure}
    \caption{Performance impacts of coarse-grained locks and more devices (1 to 32 devices).
    \emph{trylock} uses coarse-grained try locks.
    \emph{lockless} removes coarse-grained locks.}
    \label{fig:factor_study-threading1}
\end{figure}

We then studied the impact of using more devices and removing the coarse-grained lock. Figure~\ref{fig:factor_study-threading1} shows the performance result of \emph{flood\_8b} and \emph{octotiger}. \emph{lci\_d[n]} use $n$ devices without coarse-grained locks. \emph{lci\_try\_d[n]} uses $n$ devices with coarse-grained try locks. We see that using more devices can significantly improve the maximum message rate supported by the runtime in both cases, as we have more communication resources. Removing the coarse-grained locks can further improve the performance and reach the maximum message rate with fewer devices (which means less NIC resource/memory consumption and better scalability). However, these improvements do not translate into application performance because the application does not stress communication performance as much.

\obs{Thread contention on coarse-grained blocking locks significantly impacts AMTs performance. Replacing the blocking locks with try locks and allocating more resources improves application performance. Removal of the coarse-grained locks can further improve the performance at the microbenchmark level. In addition, AMTs would favor an explicit progress function: An explicit progress function enables clients to control progress frequency, which, depending on the lock strategy, can have significant effects on application performance.}

\section{Related Work}

\label{sec:related_work}
\subsection{AMTs and their Communication Layer}
Various AMTs have been proposed in recent years. We focus on the ones targeting distributed memory architecture, 
such as PaRSEC~\cite{bosilca2013PaRSECExploitingHeterogeneitya}, Legion~\cite{bauer2012LegionExpressingLocality}, HPX~\cite{Kaiser2020HPX}, StarPU~\cite{augonnet2009StarPUUnifiedPlatform}, Charm++~\cite{kale1993CHARMPortableConcurrent}, and DDDF~\cite{chatterjee2013IntegratingAsynchronousTask}. 
With such systems, users generally do not need to call external communication libraries. This simplifies programming and allows for better integration of the communication library with the task scheduler. 
PaRSEC and HPX use MPI and LCI as their communication backends. Legion has GASNet-EX~\cite{bonachea_gasnet-ex_2018}, MPI, and UCX~\cite{shamis2015ucx} backends. Charm++ supports multiple communication backends, including MPI, UCX, Libfabric, Libibverbs, uGNI~\cite{cray2019uGNI}, and PAMI~\cite{Kumar2012PAMI}. StarPU is based on MPI and NewMadeleine~\cite{Aumage2007nmad}. Most of their publications focus on the whole system and only briefly describe their communication layer design. Their experiment evaluations do not usually cover design decisions in the communication layers. This paper primarily focuses on the communication layer design of HPX and systematically studies each design decision.

Based on the literature review and code inspection, we could identify many techniques in other systems similar to those discussed here. Realm~\cite{treichler2014RealmEventbasedLowlevel} (Legion's low-level task engine), PaRSEC, and Charm++ all have a one-sided-style communication abstraction that involves small header messages for control signal and large followup messages for data transfer. Realm, PaRSEC, and Charm++ all have the concept of progress threads (or communication threads) and their worker threads typically can directly issue communications. The MPI backends of Charm++, Realm, and DDDF all use pre-posted receives for control messages and some linked list or queue to store the pending MPI requests and periodically poll them with \emph{MPI\_Test}. The MPI backend of PaRSEC uses a more sophisticated two-layer approach that uses \emph{MPI\_Testsome} to poll a small subset of prioritized pending MPI requests, but there are no published evaluations of the efficiency of this scheme. None of them use resource replication, except for a personal fork of the Legion MPI backend that briefly evaluates the MPICH multi-VCI setting~\cite{zambre2021LogicallyParallelCommunication} but the configuration is not available in the Legion master branch.

~\cite{mor2023PaRSEC_LCI} studied the integration between PaRSEC and LCI. Many design choices discussed here were not explored because PaRSEC hardcodes these choices in its upper communication layer, which limits the flexibility to refactor. For example, its upper layer funnels most communication activities to a single thread. 
While a workshop paper~\cite{yan2023design} described a work-in-progress version of the LCI parcelport with a legacy LCI version, 
this paper presents a much more optimized LCI parcelport with new techniques and many-fold performance improvement, systematic performance contribution evaluation for each technique, and insights beyond HPX and LCI.

\subsection{Related Communication Layer Optimization}

Many communication library constructs discussed here are not unique to LCI. Many similar techniques exist in other communication libraries. 
GASNet-EX~\cite{bonachea_gasnet-ex_2018} is the most established communication library focusing on irregular communication supports. It offers active messages as one of the primary communication primitives and thus is capable of direct support for the one-sided control message transfer. It also offers an explicit progress function for users to poll. 
Many lessons learned here are also relevant to GASNet-EX. As discussed earlier, we cannot directly process the received parcels in the active message handler and thus still need a high-performance MPMC queue to move the execution context outside the communication runtime. In addition, GASNet-EX still needs to improve its local completion mechanism and multithreading supports. Currently, GASNet-EX users can only use the event object to poll for each individual operation, which has the same inefficiency as the MPI request pool approach. GASNet-EX has also proposed a multi-domain optimization similar to the multi-device optimization used here but the multi-domain optimization can only be applied to its remote-memory-access (RMA) primitive but not active messages~\cite{ibrahim2014gasnet_domain}. 

The MPI community has proposed various vendor-specific extensions and new standard proposals similar to the communication requirement discussed here. The MPICH VCI extension~\cite{zambre2020vci} enables users to use multiple low-level communication resources by mapping different communicators to different low-level devices, similar to what we do with LCI devices. However, this extension lacks fine-grained control to selectively replicate resources for only a subset of MPI communicators. This limitation reduces its usability in scenarios where communicators are used for both collective communication and resource replication purposes.
There are similar extensions in OpenMPI (CRI)~\cite{Patinyasakdikul2019openmpi_mt} but they are not available in the commonly used OpenMPI UCX or OFI backends. 
The MPIX continuation proposal~\cite{schuchart2021CallbackbasedCompletionNotification} supports attaching callback functions to MPI requests. This would enable the MPI parcelport to poll for many pending operations more efficiently, similar to what the LCI parcelport does with completion queues. A high-performance MPMC queue is still needed to move the execution context outside the MPI runtime. In addition, the lack of an explicit progress function in the MPI standard necessitated the addition of a new  "continuation request" type, which adds to the complexity of the design. The reference implementation has not been merged into the OpenMPI repository. There is a new MPIX stream proposal~\cite{zhou2022mpix_stream} proposed by the MPICH team that enables explicit VCI specification and progress invocation, but it is still in an early stage. The MPI partitioned communication~\cite{grant2019FinepointsPartitionedMultithreaded} is proposed to improve MPI's threading/task support and has been included in the MPI standard since MPI 4.0. It enables different threads/tasks to contribute small data chunks to a single MPI communication. However, it does not fit into the HPX communication abstraction as HPX tasks are less collaborative and the communication logic is more dynamic.

While these existing techniques and extensions can be helpful, they are scattered across different communication libraries and thus are hard to use for a systematic investigation of their effectiveness in a common framework, as we do with the LCI parcelport.

\section{Conclusion}
\label{sec:conclusion}

We have described the HPX parcelport communication layer.
Based on the existing MPI parcelport implementation inefficiencies, we identified four areas that might benefit from better support by the communication library. We implemented such support in the LCI parcelport and evaluated the overall performance improvement and the impact of each change. Our results demonstrate that AMTs can significantly benefit from advanced features and optimizations in communication libraries. Among the four areas, better multithreading and concurrency requirements are the most important: AMTs must support efficient polling for many operations and enable concurrent communication across all layers. An explicit progress function is also favorable to ensure sufficient progress in communication.
On the other hand, providing a more direct primitive for unexpected messages, despite its potential advantage, has yet to demonstrate practical benefits. Overall, thread contention, particularly lock contention, emerges as the most crucial factor. Many of the techniques described in this paper had a significant impact because they addressed thread contention in one way or another.


\end{document}